**Theoretical Frameworks for Integrating Sustainability Factors into Institutional Investment Decision-Making**

**Innocentus Alhamis**
**Southern New Hampshire University**
**i.alhamis@snhu.edu**

**Dr. Innocentus Alhamis is a professor at the School of Business at Southern New Hampshire University.**

**Abstract**

This paper explores key theoretical frameworks instrumental in understanding the relationship between sustainability and institutional investment decisions. The study identifies and analyzes various theories, including Behavioral Finance Theory, Modern Portfolio Theory, Risk Management Theory, and others, to explain how sustainability considerations increasingly influence investment choices. By examining these frameworks, the paper highlights how investors integrate Environmental, Social, and Governance (ESG) factors to optimize financial outcomes and align with broader societal goals.

**Keywords**: Sustainability, Institutional Investors, Theoretical Frameworks

**1.0 Introduction**

This paper emphasizes the critical role of theoretical frameworks in explaining how institutional investors integrate sustainability criteria into their decision-making processes. As the investment landscape evolves, institutional investors face mounting pressure from diverse stakeholders, including individual investors, policymakers, regulatory agencies, the United Nations (UN), and other influential entities (Eccles et al., 2014). These actors increasingly advocate for incorporating environmental, social, and governance (ESG) factors alongside traditional profit objectives into investment strategies. The growing awareness of the impact of investment decisions on societal and environmental outcomes has led to a paradigm shift in how institutional investors approach their portfolios (Eccles et al., 2014; Friede et al., 2015; UN PRI, 2020; Dervi et al., 2022 and Hyrske et al., 2022)

Numerous studies have explored the integration of sustainability into the decision-making frameworks of institutional investors, highlighting the critical role of Environmental, Social, and Governance (ESG) criteria in shaping investment strategies (Eccles et al. (2014); Friede et al., 2015; UN PRI, 2020; Wang & Zhang, 2016 and Hyrske et al.,2022). Research indicates that incorporating ESG factors enhances ethical considerations and positively impacts financial performance and risk management (Eccles et al., 2014; Friede et al., 2015). By effectively blending ESG principles into their decision-making processes, investors can better align their portfolios with sustainable practices while potentially improving long-term returns (Friede et al., 2015).

Moreover, regulatory initiatives such as the UN Principles for Responsible Investment (UN PRI) further encourage institutional investors to adopt sustainable practices, thus promoting long-term value creation (UN PRI, 2020). Research has shown that integrating sustainability into investment decisions significantly enhances risk assessment, with ESG factors as indicators of a company's long-term viability and resilience. For example, firms that maintain strong environmental practices are often better positioned to avoid regulatory penalties and mitigate reputational risks. Additionally, organizations prioritizing social responsibility typically experience greater employee engagement and customer loyalty, which can stabilize their market position (Khan et al., 2016; Eccles et al., 2014; Dervi et al., 2022).

Integrating Environmental, Social, and Governance (ESG) criteria into investment strategies can enhance portfolio diversification by broadening the range of assets and mitigating concentration risks. Studies show that ESG-focused portfolios tend to exhibit lower volatility and improved risk-adjusted returns over time, underscoring the potential for ESG integration to contribute to more resilient investment portfolios (Krueger et al., 2020; Wang et al., 2016; Khan et al., 2016). Thus, by integrating ESG factors, investors gain access to a broader range of assets, potentially reducing risk and fostering a more resilient investment portfolio.

The increasing focus on sustainability and ESG criteria transforms institutional investing by fostering a more responsible and socially conscious approach to decision-making. This shift aligns financial goals with broader societal values, addressing the demand for accountability among investors while promoting sustainable economic development and responsible corporate practices. By incorporating ESG factors, institutions can support long-term value creation that resonates with financial and ethical priorities (Khan et al., 2016; Eccles et al., 2014; Dervi et al., 2022).

Overall, the literature underscores the significance of ESG integration as essential for fostering ethical investment practices and improving financial outcomes. By incorporating Environmental, Social, and Governance criteria, investors are better positioned to align with societal values while achieving stronger risk-adjusted returns and enhanced portfolio resilience.

## 2.0 Purpose of the study

While existing literature has extensively discussed how sustainability is integrated into decision-making processes, there remains a significant gap in research specifically addressing the theoretical frameworks that support the incorporation of Environmental, Social, and Governance (ESG) factors into the investment decisions of institutional investors. This paper seeks to bridge this gap by offering a comprehensive exploration of the foundational theories that underpin the integration of ESG considerations into investment strategies. By examining relevant theoretical frameworks, the study aims to enhance our understanding of how sustainability influences investment decision-making, aligning financial objectives with broader societal and environmental goals. Through this exploration, the research contributes to the growing discourse on incorporating sustainability into institutional investment practices, an increasingly vital area in today's financial landscape.

## 3.0 Theoretical Framework

A theoretical framework is a structured set of concepts, definitions, and propositions that facilitates the analysis of a specific phenomenon (Creswell, 2014). A theoretical framework is essential for guiding research and practice across various disciplines. Such frameworks provide structured lenses through which complex phenomena can be analyzed and understood. They help clarify concepts, define variables, and delineate their relationships, enabling researchers to develop coherent hypotheses and methodologies (Creswell & Creswell, 2017; Maxwell, 2013).

Moreover, theoretical frameworks can enhance the credibility and reliability of research findings by providing a consistent basis for interpretation and evaluation. They help identify gaps in existing knowledge and stimulate further investigation, ultimately contributing to the advancement of theory and practice in sustainability and investment (Ravitch & Riggan, 2016).

By employing a theoretical framework, researchers can move beyond simple description, enabling the formulation of generalizations about a phenomenon and its underlying factors (Ravitch & Riggan, 2016). Additionally, theoretical frameworks help define these generalizations' boundaries, clarifying the scope and limitations of a study's findings (Yin, 2018). Theoretically, theoretical frameworks provide a fundamental basis for understanding and interpreting research results by specifying the variables most significantly influencing a phenomenon.

## 4.0 Theoretical Framework, Sustainability, and Institutional Investors' Decisions

In the context of sustainability and investment, theoretical frameworks are particularly crucial. They allow investors and researchers to navigate the intricate interactions between environmental, social, and governance (ESG) factors and financial performance. Theoretical frameworks provide a structured approach to understanding how environmental, social, and governance (ESG) considerations influence institutional investment decisions. They guide investment strategies, improve risk management practices, evaluate performance impacts, influence corporate behavior, enhance market efficiency, and support regulatory developments in ESG investing (Clark et al., 2004; Eccles & Serafeim, 2013).

The integration of sustainability considerations in institutional investment decisions is underpinned by diverse theoretical perspectives that underscore the significance of ESG factors in assessing risk, enhancing long-term value creation, and aligning investments with societal expectations. By adopting these frameworks, institutional investors can contribute to sustainable development goals while pursuing financial objectives, promoting a more resilient and responsible investment landscape (Clark et al., 2004; Eccles & Serafeim, 2013).

Theoretical frameworks establish a robust foundation for understanding the rationale behind institutional investors' adoption of sustainability considerations in their investment strategies. They shape corporate behavior and contribute to fostering a sustainable global economy (Clark et al., 2004; Eccles & Serafeim, 2013). As such, theoretical frameworks provide essential insights into how institutional investors can navigate this complex landscape, balancing profitability with responsible investment practices (Eccles & Klimenko, 2019).

## 5.0 Theoretical Overview

This study has highlighted several key theories essential to understanding sustainability and institutional investment decisions. These include Behavioral Finance Theory, Modern Portfolio Theory, Risk Management Theory, Value Creation Theory, Institutional Theory, Agency Theory, Socially Responsible Investment Theory, and Reputational Theory.  Each of these theories provides a unique lens for understanding how sustainability considerations can be integrated into institutional investment strategies, supporting financial and ethical goals.

### 5.1 Behavioral Finance Theory

Behavioral finance theory is a field that explores the psychological factors influencing the financial behaviors of individuals and institutions (Thaler,1980; Tversky & Kahneman, 1974; Barberis & Thaler, 2003; Morvan et al., 2017). Unlike traditional finance, which operates under the assumption that investors are rational and always act to maximize utility, behavioral finance recognizes that human behavior often deviates from rational decision-making due to various cognitive biases and emotional influences (Shiller, 2000; Mody, 2018).

Behavioral finance focuses on the influence of cognitive biases—systematic patterns of thinking that lead to irrational judgments—on investors' decision-making. For instance, overconfidence can cause investors to overestimate their knowledge or predictive abilities, often leading them to take on more risk than is warranted (Shiller, 2000). Another significant bias is anchoring, where individuals rely too heavily on the first piece of information they encounter when making decisions, potentially leading to poor investment choices (Tversky & Kahneman, 1974; Morvan et al., 2017).

Moreover, emotional factors are crucial in investment decisions (Kshneman, 2011; Tversky & Kahneman, 1974; Shiller, 2000; Mody, 2018). Emotions like fear and greed can drive irrational market behaviors, contributing to market bubbles and crashes (Shiller, 2000). Behavioral finance examines how these emotional responses manifest in trading behaviors, impacting investors' decision-making processes (Tversky & Kahneman, 1974; Morvan et al., 2017; Mody, 2018).

Heuristics, or mental shortcuts, are another important aspect of behavioral finance. While these heuristics can help simplify complex decisions, they may also lead to suboptimal investment outcomes. For instance, representativeness heuristic can cause investors to make judgments based on how closely a situation resembles a typical case, often overlooking relevant statistical data (Tversky & Kahneman, 1974).

Behavioral finance also explains market anomalies that traditional financial theories struggle to account for. Examples include the equity premium puzzle, which refers to the larger-than-expected historical returns of stocks over bonds, and the January effect, where stock prices tend to rise in January due to various psychological factors (Fama, 1998).

Investors and financial professionals must comprehend behavioral insights influencing sustainability and institutional investment decisions. By recognizing the cognitive biases and emotional influences at play, they can develop strategies that help mitigate these effects, ultimately promoting better investment decision-making.

This framework is particularly relevant when examining sustainability and investment decisions, as it helps to understand why investors may favor or disregard sustainable practices. Investors often exhibit biases such as overconfidence, anchoring, and herd behavior, leading to irrational decision-making. For instance, overconfidence can result in underestimating risks associated with investments in non-sustainable companies, while herd behavior may cause investors to flock toward popular but potentially unsustainable trends (Barberis & Thaler, 2003). These biases may hinder the integration of sustainability considerations into investment strategies, as investors might prioritize short-term gains over long-term sustainability outcomes.

Behavioral finance also helps to explain the growing interest in sustainable investing. As concerns about climate change and social issues increase, shifting societal values influence investor preferences, driving demand for sustainable and ethical investment options. The rise of environmental, social, and governance (ESG) criteria reflects a collective behavioral change where investors increasingly seek to align their portfolios with ethical values (Eccles et al., 2014; Khan et al., 2016). This shift suggests that investors are responding to new social pressures, which can create a market for sustainable investments.

Additionally, the framing of sustainability-related information can significantly impact investment decisions. How information about sustainable practices is presented can either attract or deter investors. For example, emphasizing potential long-term financial benefits from sustainable investments may counteract initial biases toward traditional investment options (Eccles et al., 2014).

In summary, the interplay between behavioral finance and sustainability highlights how cognitive biases can obstruct rational investment decisions while showing how changing norms can encourage sustainable investing. Understanding these dynamics is crucial for financial advisors and firms promoting sustainable investment strategies.

### 5.2 Modern Portfolio Theory

Modern Portfolio Theory (MPT), introduced by Harry Markowitz in the early 1950s, laid the groundwork for contemporary investment strategies by emphasizing the importance of diversification, risk assessment, and the optimization of investment portfolios to achieve desired financial outcomes (Markowitz, 1952). The theory provides a framework for constructing an investment portfolio that aims to maximize the expected return based on a given level of risk.

According to this theory, investors are risk-averse and seek to optimize their portfolios by diversifying their investments across various asset classes (Markowitz, 1952).

The theory posits that a portfolio's overall risk is shaped not merely by the risks of individual assets but by the way these assets relate to one another. This relationship, measured by the correlation of asset returns, is central to constructing diversified portfolios. Through diversification, investors can achieve an "efficient frontier," representing the set of portfolios offering the maximum expected return for a given level of risk (Markowitz, 1952; Elton & Gruber, 1997). This concept emphasizes that investors can reduce risk and potentially enhance returns by selecting a mix of assets with low or negative correlations.

The theory also introduced the idea of the mean-variance optimization framework, where investors can calculate a portfolio's expected return and risk (standard deviation) based on the expected returns of the individual assets, their variances, and their covariances (Markowitz, 1952). It emphasizes the importance of asset allocation and diversification in managing risk, suggesting that a well-constructed portfolio can achieve better risk-adjusted returns than individual assets alone (Fama & French, 2004; Sharpe, 1964).

The Capital Asset Pricing Model (CAPM), developed in the 1960s, extended MPT by establishing a linear relationship between the expected return of an asset and its systematic risk (Mossin, 1966). CAPM helps investors understand how the risk of an asset, in relation to the market, influences its expected return, thereby guiding investment decisions (Sharpe, 1964; Lintner, 1965).

In the context of sustainability and investors' decision-making, Modern Portfolio Theory (MPT) can be linked to sustainability by incorporating Environmental, Social, and Governance (ESG) factors as additional determinants of risk and return. The theory explains how investors can construct financially optimal portfolios aligned with sustainable and responsible investing objectives.

Integrating ESG considerations into MPT-based strategies provides a balanced means for investors to achieve sustainable growth while managing risks linked to environmental, social, and governance issues. In MPT, risk and return are not only a function of individual assets but of the correlations between them. Similarly, ESG criteria can be considered additional risk and potential return dimensions, expanding MPT's traditional financial framework. Incorporating ESG factors into portfolio construction, investors can create portfolios aligning with financial and sustainability objectives. This approach allows for an "efficient frontier" that meets expected returns for given levels of risk and aligns with ethical and sustainable investment principles, enabling investors to address financial and social priorities (Markowitz, 1952; Friede et al., 2015).

Furthermore, ESG factors add a layer of risk management by identifying companies facing regulatory challenges, reputational damage, or operational disruptions due to unsustainable practices. For example, companies that adopt strong environmental standards may reduce risks associated with climate change regulations and resource scarcity, making them more resilient in the long term. Studies have shown that integrating ESG considerations into portfolios can help mitigate idiosyncratic risks and yield more stable returns, thus contributing to better risk-adjusted performance (Friede et al., 2015; Eccles et al., 2014).

Incorporating sustainability also aligns with changing investor preferences, as many now seek portfolios that reflect ethical and social values. Sustainable investment decisions often use ESG metrics to screen out companies with poor social or environmental records while including firms with strong sustainability practices (Friede et al., 2015; Khan et al., 2016). This approach aligns with investors' ethical goals and can provide financial advantages by targeting companies better positioned for long-term growth. Wang et al.

(2016) found that portfolios with ESG-focused companies exhibited lower volatility and better diversification benefits, aligning well with MPT's diversification principles.

Moreover, regulatory shifts, such as the UN Principles for Responsible Investment (PRI), encourage institutions to integrate sustainability as a core investment principle, promoting longer-term value creation. For investors, using ESG criteria within MPT allows for alignment with global sustainability standards and enhances portfolio resilience while meeting growing demands for responsible investments (PRI, 2020).

### 5.3 Risk Management Theory

Risk management theory is a foundational concept in finance and business that focuses on identifying, assessing, and mitigating risks to achieve desired outcomes (Merna et al., 2012; Friede et al., 2015; Hull et al., 2018). Following this process, aligned efforts are made to minimize, monitor, and control the likelihood or impact of adverse events (Beasley et al., 2005). This theory offers a framework for analyzing how different types of risks, including both financial and non-financial factors, influence investment results.

In sustainability and institutional investment decisions, robust risk management is essential for identifying, assessing, and mitigating various Environmental, Social, and Governance (ESG) risks. Effectively managing these risks allows investors to safeguard against potential adverse outcomes while seeking sustainable, long-term returns. By integrating ESG factors into risk management frameworks, institutional investors can better align their portfolios with financial goals and responsible investing practices (Eccles et al., 2014; Khan et al., 2016). This approach helps to minimize exposure to risks that may arise from environmental degradation, social unrest, or governance failures, ultimately enhancing the resilience and sustainability of investment portfolios. Studies show that companies with strong risk management practices criteria often demonstrate lower risk profiles and tend to perform better financially over time (Eccles et al., 2014; Khan et al., 2016).

ESG risks are typically classified into three primary categories: environmental, social, and governance. Environmental risks include climate change, resource scarcity, and pollution. Companies that do not adequately address these issues may face regulatory penalties, operational disruptions, and reputational damage. For instance, extreme weather events can disrupt supply chains, leading to financial losses (Eccles et al., 2014).

Social risks pertain to how a company manages relationships with employees, customers, and communities (Friede et al., 2015; Khan et al., 2016). Poor labor practices, inadequate health and safety measures, and negative community impacts can lead to significant reputational and legal risks. Research shows that firms neglecting social responsibility may experience volatility in stock prices due to public backlash (Dimson et al., 2015).

Governance risks relate to a company's leadership, board structure, and ethical conduct. Weak governance practices can lead to mismanagement, fraud, and diminished shareholder value. Strong governance frameworks are correlated with better risk management and financial performance (Gibson et al., 2019).

Investors can use quantitative methods to measure ESG risks, often employing statistical models to analyze the correlation between ESG factors and financial performance. This data-driven approach helps understand the potential financial impact of various ESG risks on investment returns (Khan et al., 2016).

In addition to quantitative methods, qualitative assessments involve analyzing a company's ESG policies, stakeholder engagement, and industry practices. This holistic view allows investors to gauge how well a company manages its ESG risks and identify improvement areas (Friede et al., 2015).

One of the most effective strategies for managing ESG risks is diversification. By spreading investments across various sectors and asset classes, investors can reduce the overall risk exposure associated with any single investment. This approach mitigates the impact of adverse events related to ESG factors (Markowitz, 1952).

Investors increasingly view ESG factors as essential components of risk assessment. This theory highlights that companies with robust ESG practices are often better equipped to handle regulatory changes, reputational risks, and operational challenges (Eccles et al., 2014; Khan et al., 2016). By integrating ESG into their analysis, investors can identify potential risks that traditional financial metrics may overlook, ultimately leading to more resilient investment portfolios.

### 5.4 Value Creation Theory

Value creation theory focuses on how businesses generate value for stakeholders, including shareholders, employees, customers, and society. This theory is essential for understanding the role of a business in contributing positively to the economy, environment, and social well-being. Value creation involves leveraging resources, capabilities, and competitive advantages to produce value that benefits all stakeholders (Freeman et al., 2004; Lepak et al., 2007; Porter et al., 2011; Grönroos et al., 2013 and Fichtenbauer, 2015)

Traditionally, value creation was defined mainly in financial terms, where maximizing shareholder wealth was the primary objective (Fama, 1970). However, contemporary views of value creation emphasize sustainable and ethical practices that support long-term societal and environmental health. This approach to value creation is particularly aligned with corporate social responsibility (CSR) and environmental, social, and governance (ESG) initiatives, which demonstrate that sustainable practices can drive financial outperformance by building trust and fostering loyalty (Clark et al., 2014; Eccles et al., 2014; Joudi et al., 2024)

Value creation theory closely aligns sustainability in driving investment decisions, as it emphasizes building long-term value that benefits not only shareholders but all stakeholders, including employees, communities, and the environment (Freeman, 2004). By integrating environmental, social, and governance (ESG) factors into their business strategies, companies aim to meet sustainable development goals while managing risks associated with environmental and social issues, which can have direct financial impacts.

Sustainability has become increasingly important in value creation, especially as evidence suggests sustainable practices can lead to stronger financial performance and resilience. This is due to enhanced operational efficiencies, reputational gains, and stronger relationships with customers and communities, which, in turn, influence investor confidence. For instance, Eccles, Ioannou, and Serafeim (2014) found that companies prioritizing corporate sustainability showed improved operational and stock performance, suggesting that sustainability can be a valuable driver of corporate success.

Investors today often evaluate companies based on ESG criteria, which offer insights into a company's commitment to sustainable practices and long-term value. By incorporating ESG factors, investors can identify companies that manage risks better and foster innovations that align with societal trends and regulations. As Khan, Serafeim, and Yoon (2016) point out, companies addressing material sustainability issues tend to outperform in areas critical to their industries, making them attractive investment opportunities for institutional and individual investors. Thus, value creation through sustainability can be a decisive factor in investment decisions, as it supports both financial and societal returns, making it integral for investors seeking growth and responsible investment.

From a theoretical perspective, value creation can be seen as aligning business strategy with stakeholder theory, recognizing that a company's success relies on fulfilling the needs of diverse stakeholders. By integrating CSR and ESG into core strategies, companies can achieve competitive advantages and enhance brand reputation, ultimately contributing to sustainable financial returns (Khan et al., 2016).

This theory posits that companies engaging in sustainable practices are more likely to generate long-term value. By considering ESG factors, investors can identify firms committed to sustainability and social responsibility, which can lead to enhanced financial performance over time. This aligns with the growing recognition that ESG investing is a moral and strategic choice.

### 5.5 Institutional Theory

Institutional theory is a framework that examines how institutions—defined as rules, norms, and beliefs—shape social behavior and organizational structures (Meyer et al.,1977; DiMaggio & Powell, 1983; Thornton et al., 2012). It emphasizes the significance of social and cultural contexts in influencing the practices and decisions of organizations. The theory posits that organizations are driven by market forces and the need to gain legitimacy within their environments.

DiMaggio and Powell (1983) introduced the concept of institutional isomorphism, which describes how organizations become similar over time due to coercive, mimetic, and normative pressures. Coercive isomorphism arises from formal and informal pressures from other organizations or societal expectations. Mimetic isomorphism occurs when organizations model themselves after others, especially during times of uncertainty. Normative isomorphism relates to professionalization and the influence of norms and values within a field.

Furthermore, Scott (2014) expanded on institutional theory by identifying three pillars: regulatory, normative, and cultural-cognitive. The regulatory pillar encompasses laws and regulations that institutions impose. The normative pillar consists of values and norms that shape the behaviors of individuals and organizations. In contrast, the cultural-cognitive pillar refers to shared beliefs that create meaning within a society.

Overall, institutional theory highlights the complexity of organizational behavior and the importance of understanding the broader societal and institutional context in which organizations operate (Meyer & Rowan, 1977; North, 1990).

Institutional theory provides a valuable framework for understanding how external pressures, such as regulations, societal norms, and industry standards, influence institutional investors' decisions regarding sustainability and institutional investment decisions.

Regulatory pressures are a significant driver of corporate sustainability. Governments and regulatory bodies increasingly mandate that companies adhere to environmental standards and disclose their sustainability practices. Such regulations can compel organizations to integrate sustainable practices into their operations, as failure to comply can result in legal repercussions and reputational damage. This aspect of institutional theory highlights how coercive pressures can lead organizations to adopt more sustainable practices, making them more appealing to investors concerned about risks associated with regulatory non-compliance (Thornton et al., 2012; Scott, 2014).

Normative pressures also play a crucial role in this relationship. Organizations often feel the need to conform to the norms and values prevalent in their industry or community. For instance, firms that embrace sustainability enhance their legitimacy in the eyes of stakeholders and align themselves with broader societal expectations. This alignment is increasingly important to investors seeking opportunities in firms committed to social and environmental responsibility. DiMaggio and Powell (1983) describe this phenomenon of isomorphism, where companies mimic the successful sustainability practices of their peers to gain legitimacy. As more investors incorporate Environmental, Social, and Governance (ESG) criteria into their decision-making, the market rewards companies that visibly engage in sustainable practices (Eccles et al., 2014).

The cultural-cognitive dimension of institutional theory further illustrates how shared beliefs about sustainability can influence investor behavior. As sustainability becomes a fundamental component of corporate strategy and culture, it reshapes investor expectations and preferences. Investors are increasingly inclined to support companies that comply with regulations and demonstrate a proactive commitment to sustainability. This shift reflects a growing recognition of the long-term value that sustainable business practices can deliver (Freeman, 1984; Friede et al., 2015).

By examining the role of these external forces, institutional theory helps explain how sustainability considerations are integrated into investment practices. As institutional investors face increasing demands for accountability and alignment with socially responsible investment practices, sustainability becomes a crucial factor in their decision-making processes (DiMaggio & Powell, 1983; Scott, 2014)

In summary, institutional theory helps elucidate how external pressures and internal motivations within organizations shape sustainability practices. As companies navigate these institutional environments, their sustainability initiatives influence investor decisions. Investors, recognizing the financial and reputational benefits of sustainability, increasingly favor firms that align with these values, leading to a convergence of corporate practices with investor expectations.

### 5.6 Agency Theory

Agency theory is a framework that examines the relationship between principals (such as shareholders) and agents (such as company executives) in business settings. It is grounded on the premise that the interests of these two parties may diverge, leading to potential conflicts of interest (Fama & Jensen, 1983; Alsharqawi et al., 2019). This theory primarily addresses the issues arising from the principal-agent relationship, particularly when agents are tasked with making decisions on behalf of principals.

The foundational idea behind agency theory is that principals delegate authority to agents to act on their behalf. However, agents may not always act in the best interests of the principals (Jensen & Meckling, 1976). This misalignment can lead to agency costs, which are the costs incurred by the principal to monitor the agent's actions, ensuring that they align with the principal's objectives (Jensen & Meckling, 1976; Alsharqawi et al., 2019). These costs can take the form of monitoring expenses, bonding costs incurred by agents to guarantee they act in the principal's best interests, and residual loss, representing the loss of value from a divergence of interests.

To mitigate agency problems, principals often employ various governance mechanisms. These include performance-based compensation to align the interests of agents with those of principals, increased transparency, and the establishment of boards of directors to oversee management activities (Fama & Jensen, 1983). However, while these mechanisms can help reduce agency costs, they cannot eliminate them entirely.

Agency theory has been instrumental in shaping corporate governance practices and understanding the dynamics between shareholders and management. It provides valuable insights into executive compensation, managerial incentives, and the importance of regulatory frameworks that promote organizational transparency and accountability (Eisenhardt, 1989; Alsharqawi et al., 2019).

Agency theory is particularly relevant to sustainability and investors' investment decisions because it highlights conflicts between management (agents) and shareholders (principals) regarding long-term sustainability practices.

As companies increasingly prioritize sustainability, the tension between short-term profit maximization and long-term sustainability initiatives becomes more pronounced. Agency theory suggests that managers may prioritize their own interests—such as receiving bonuses linked to short-term financial performance—over long-term sustainability goals that could benefit shareholders in the future (Jensen & Meckling, 1976). This misalignment can lead to agency costs, as shareholders might need to implement monitoring mechanisms to ensure that management commits resources to sustainable practices rather than merely focusing on immediate financial returns (Fama & Jensen, 1983).

Furthermore, the rise of socially responsible investing (SRI) and environmental, social, and governance (ESG) criteria has increasingly prompted investors to consider sustainability in their investment decisions. Investors may demand that management adopt sustainable practices, viewing them as essential for long-term financial performance and risk mitigation. Research has shown that companies that engage in sustainable practices often outperform their peers in the long run, making these practices appealing to investors (Friede et al., 2015).

To address these issues, companies can employ strategies that align management incentives with sustainability goals, such as linking executive compensation to long-term ESG performance metrics. This aligns the interests of agents with those of principals, encouraging a focus on sustainability that can enhance long-term shareholder value (Eccles et al., 2014). By doing so, firms can effectively bridge the gap between short-term financial objectives and long-term sustainable practices, thus fostering a more sustainable investment landscape.

### 5.7 **Socially responsible investment (SRI)**

Socially responsible Investment (SRI) is an investment strategy incorporating environmental, social, and governance (ESG) criteria alongside traditional financial analysis. The underlying theory of SRI posits that investors can achieve competitive financial returns while also supporting positive social and environmental outcomes (Guerard, 2003; Sparkes & Cowton, 2004; Camilleri, 2017 and Baker & Nofsinger, 2012 and Ransome & Sampford, 2010)

SRI is grounded in the belief that investments should align with investors' values and ethical considerations. This includes avoiding companies involved in harmful practices such as tobacco production, weapons manufacturing, or environmental degradation (Camilleri, 2017 & Jianghong, 2023)

SRI often involves "positive screening," where investors seek to invest in companies that actively promote social good. This may include firms prioritizing renewable energy, diversity and inclusion, fair labor practices, and community engagement (Camilleri, 2017; Jianghong, 2023).

Conversely, SRI may employ "negative screening," where investors exclude certain sectors or companies based on specific ethical criteria. For example, investors might avoid companies that are involved in fossil fuels or those that have poor labor practices (Camilleri, 2017; Jianghong, 2023).

SRI emphasizes long-term value creation rather than short-term gains. Investors believe that companies committed to sustainable practices are likely to perform better over time as they are better equipped to manage risks related to environmental and social issues (Guerard, 2003; Sparkes & Cowton, 2004; Ransome & Sampford, 2010).

SRI encourages engagement with companies to promote better practices. This may involve shareholder advocacy, dialogue with management, and participation in corporate governance to encourage more responsible behavior.

An important aspect of SRI is measuring the impact of investments on social and environmental outcomes. Investors seek to assess how their portfolios contribute to sustainability and social equity.

ESG is fundamentally based on the SRI framework. Integrating Socially Responsible Investment (SRI) into Environmental, Social, and Governance (ESG) criteria represents a synergistic approach that enhances investment decision-making. This integration helps investors align their portfolios with ethical values while focusing on financial performance and sustainability.

Institutional investors are often focused on long-term growth and sustainability. SRI promotes the idea that companies with strong ESG practices are more likely to be resilient and successful over time. By investing in these companies, institutional investors can pursue strategies aligning with their long-term goals while supporting sustainable practices.

In summary, the theory of Socially Responsible Investment (SRI) advocates an investment approach that integrates ethical considerations and ESG criteria with traditional financial analysis. By prioritizing long-term value and stakeholder engagement, SRI aims to positively impact society and the environment while still achieving financial returns.

SRI can explain institutional investors' decision-making processes by providing a framework that integrates ethical considerations, risk management, long-term value creation, stakeholder engagement, and responsiveness to regulatory trends. By adopting SRI principles, institutional investors can effectively balance their financial objectives with their social responsibility and sustainability commitment.

### 5.8 Reputational Theory

Reputational theory focuses on how a firm's reputation affects its behavior, performance, and stakeholder relationships. It posits that a strong reputation can be a critical intangible asset, influencing customer loyalty, employee satisfaction, and investor confidence. A positive reputation often results from consistent ethical behavior, high-quality products or services, and effective communication strategies, making it essential for firms to manage their reputations actively (Fombrun, 1996; Helm, 2007; Gibson et al., 2019; Schwaiger, 2004).

One core aspect of reputational theory is the "reputation risk," which refers to the potential loss a company might face due to negative perceptions or incidents that tarnish its reputation. Organizations invest significant resources in reputation management to mitigate these risks, as a strong reputation can lead to competitive advantages, such as customer preference and greater market share (Akerlof, 1970). Conversely, firms with a damaged reputation may struggle to recover, facing challenges in attracting customers and investors (Mishina et al., 2010; Schwaiger, 2004; Qonita et al., 2022).

Moreover, reputational theory is increasingly relevant in sustainability, where stakeholders demand transparency and accountability in corporate practices. Companies recognized for their commitment to environmental, social, and governance (ESG) principles can enhance their reputations and differentiate themselves in the marketplace. Research has shown that firms with strong sustainability practices enjoy better financial performance and lower capital costs (Eccles et al., 2014).

A strong corporate reputation, particularly in sustainability practices, can substantially influence investor behavior and perceptions. A company's reputation for sustainability can attract investors who prioritize environmental, social, and governance (ESG) criteria in their investment decisions. Research indicates that firms recognized for their sustainability initiatives tend to experience enhanced investor trust, leading to a broader and more loyal investor base (Eccles et al., 2014). Such firms are often perceived as lower-risk investments, as they are seen to be better at managing risks associated with environmental regulations, social unrest, and governance failures (Gibson, 2019). Consequently, these companies may enjoy lower capital costs and a higher market valuation than their less sustainable counterparts.

Additionally, negative events that threaten a firm's reputation can lead to immediate and lasting impacts on investor confidence. For example, companies that face scandals or fail to meet sustainability expectations can see a sharp decline in stock prices as investors react to perceived risks associated with poor reputation management (Mishina et al., 2010). This illustrates how investors increasingly make decisions based on firms' reputational capital, assessing current performance and potential risks.

Moreover, effective reputation management concerning sustainability can enhance stakeholder engagement and improve overall corporate performance. Firms that communicate their sustainability efforts transparently and authentically will likely foster strong relationships with stakeholders, including investors (Du, Bhattacharya, & Sen, 2010). This engagement can lead to more favorable investment decisions and long-term financial success.

In summary, reputational theory shapes the relationship between sustainability and investors' decisions. A robust reputation for sustainability not only attracts investment but also serves as a buffer against risks associated with negative perceptions, ultimately influencing the financial performance of firms (Eccles et al., 2014).

Table 1 below presents a concise summary of the primary focus of each theory, highlighting their key areas of exploration.

| Theory | Main Focus | Key Sources |
|---|---|---|
| Behavioral Finance Theory | examines how psychological factors influence investor behavior | (Kahneman & Tversky, 1979; Shiller, 2000; Barberis & Thaler, 2003; Morvan et al., 2017; Mody, 2018) |
| Modern Portfolio Theory | Centers on optimizing risk and return in portfolio selection | (Markowitz, 1952; Sharpe, 1964; Lintner, 1965; Mossin, 1966; Elton & Gruber,1997; Fama & Frnch, 2004) |
| Risk Management Theory | addresses the identification and mitigation of investment risks | (Beasley et al., 2005; Merna et al., 2012 and Friede et al., 2015; Hull, 2018) |
| Value Creation Theory | Investigate how sustainable practices can enhance financial performance. | (Freeman et al., 2004; Lepak et al., 2007; Porter et al., 2011; Gronroos et al., 2013; Fichtenbauer, 2015) |
| Institutional Theory | considers the role of external pressures in shaping corporate sustainability practices | (DiMaggio & Powell, 1983; Scott, 2014) |
| Agency Theory | Focus on the dynamics between investors and managers in decision-making. | (Jensen, & Meckling, 1976; Fama & Jensen,1983; Eisenhardt, 1989; Jensen, & Meckling, 2004 and Alsharqawi et al., 2019) |
| Socially Responsible Investment Theory | Promotes ethical and sustainable investment choices. | (Guerard, 2003; Sparkes & Cowton, 2004; Ransome & Sampford, 2010; Camilleri, 2017; Riedl & Smeets, 2017; Jianhong, 2023) |
| Reputational Theory | underscores the impact of a company's sustainability practices on its public perception and long-term success | (Fombrun, 1996; Schwaiger, 2004; Godfrey, 2005; Helm, 2007; Mishna et al., 2010; Qonita et al., 2022) |

Table 1: Overview of Theories and Their Main Focus

## 6.0 Conclusion

The theories examined in this study—Behavioral Finance Theory, Modern Portfolio Theory, Risk Management Theory, Value Creation Theory, Institutional Theory, Agency Theory, Socially Responsible Investment Theory, and Reputational Theory—offer critical insights into sustainability and institutional investment decisions. Each theory provides a unique perspective on how Environmental, Social, and Governance (ESG) factors influence the decision-making processes of institutional investors. By incorporating these theories, investors can align their strategies with both financial goals and sustainability objectives. Together, these frameworks highlight the importance of integrating sustainability into investment practices, enabling investors to achieve better risk management, financial returns, and ethical alignment in a rapidly evolving investment landscape.